\documentclass[referee]{raa}            

\usepackage{graphicx,times}             

\begin{document}

\title{Spectroscopic and photometric observations of Symbiotic Nova PU~Vul during 2009-2016}

   \volnopage{Vol.0 (20xx) No.0, 000--000}      
   \setcounter{page}{1}          

   \author{
       Tatarnikova A.A.\inst{1}, 
       Burlak M.A.\inst{1}, 
       Kolotilov E.A.\inst{1}, 
       Metlova N.V.\inst{1}, 
       Shenavrin V.I.\inst{1}, 
       Shugarov S.Yu.\inst{1,2}, 
       Tarasova T.N.\inst{3}, 
       Tatarnikov A.M.\inst{1} 
}

  \institute{Sternberg Astronomical Institute, Moscow State University, Moscow 119234,
Russia; {\it aat@sai.msu.ru}\\
        \and
             Astronomical Institute of the Slovak Academy of Sciences, 
             05960 Tatranska Lomnica, Slovakia\\
	\and
             Federal State Budget Scientific Institution "Crimean Astrophysical Observatory of RAS", Nauchny,
             298409, Republic of Crimea\\
    }

   \date{Received~~2017 November 30; accepted~~2017~~month day}

\abstract{
A new set of low-resolution spectral and {\it UBVJHKL}-photometric observations of the symbiotic nova PU~Vul is presented. 
The binary has been still evolving after the symbiotic nova outburst in 1977 and now it's in the nebular stage. 
It is found that the third orbital cycle (after 1977) was characterized by great changes in light curves. 
Now PU~Vul demonstrates a sine-wave shape of all light curves (with an amplitude in the {\it U} band of about 0.7 mag), 
which is typical for symbiotic stars in quiescent state. Brightness variability due to cool component pulsations is now 
clearly visible in the {\it VRI} light curves. The amplitude of the pulsations increases from 0.5 mag in {\it V} band to 0.8 mag in {\it I} band. These two types of variability, as well as a very slow change of the hot component physical parameters due to evolution 
after the outburst of 1979, influence the spectral energy distribution (SED) of the system. The emission lines variability is 
highly complex. Only hydrogen lines fluxes vary with orbital phase. An important feature of the third orbital cycle is the 
first appearance of the OVI, 6828 {\AA} Raman scattering line. We determined the hot component temperature by means of Zanstra method applied to the He II, 4686 line. Our estimate is about 150000 K for the spectrum obtained near orbital maximum in 2014. 
The VO spectral index derived near pulsation minimum corresponds to M6 spectral class for the cool component of PU~Vul.
}

\authorrunning{Tatarnikova A. A. et al. }            
   \titlerunning{Spectroscopic and photometric observations of Symbiotic Nova PU Vul during 2009-2016}  

   \maketitle

%
\section{Introduction}           
\label{sect:intro}

The symbiotic star PU~Vul (also known as the Kuwano-Honda object) is a well-known member of the small group of symbiotic novae. It means that this system has demonstrated only one outburst. And the outburst has been very slow and bright. The star exploded at the end of 1977. Near the outburst maximum, according to optical and UV data, the system was in the A-F supergiant phase and then, after a short transition period, entered the nebular phase. So, within about 10 years (1977-1988) PU~Vul, according to optical and UV spectroscopic and photometric observations, may be classified as a single red giant (before the outburst), an A-F supergiant, a Wolf-Rayet star and at last as a hot subdwarf surrounded by a gaseous nebula. Detailed bibliography concerning observations and related analysis can be found in the review of Gershberg 2000. More recent theoretical investigations of the PU~Vul outburst were published in Kato et al. 2012.

The nature of the sharp and deep brightness minimum observed in 1980 (Min. I in fig.1) had been actively debated for a long time. Spectral observations in the far UV range obtained with International Ultraviolet Explorer (IUE) discovered the eclipse nature of this minimum as well as that of the next one (minimum II), though not so sharp and deep, observed in 1993. Thus, the symbiotic nova PU~Vul is an eclipsing variable star with a very long orbital period $P\approx 13.4$ years (Shugarov et al. 2012). In this paper we consider spectral and photometrical characteristics of PU~Vul during a time interval near and after minimum III (2007).

\section{Observations}
\label{sect:Obs}

   \begin{figure}
   \centering
   \includegraphics[width=\textwidth, angle=0]{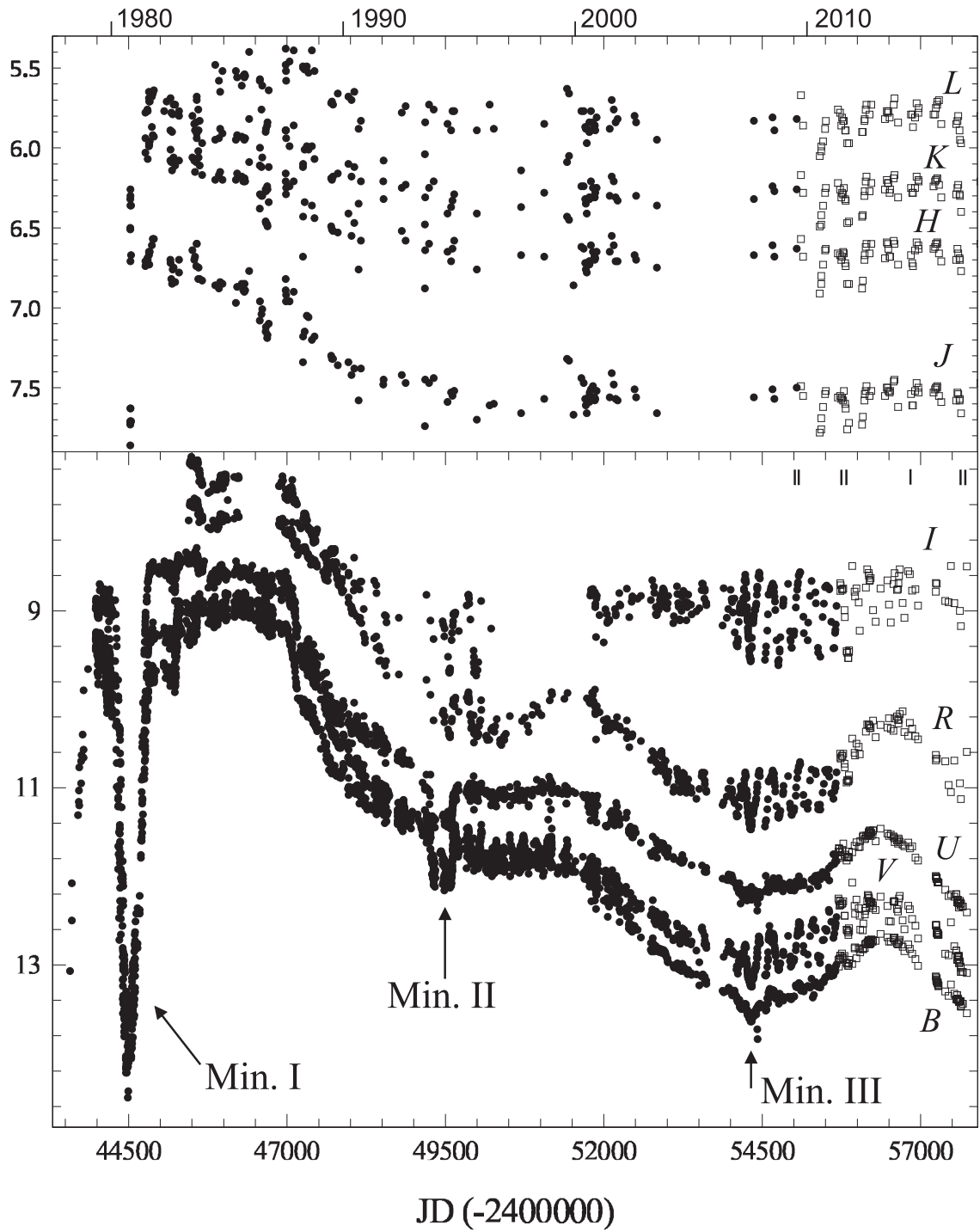}
   \caption{Light curves of the symbiotic nova PU Vul in 1977-2016. 
               Points are the data from Shugarov et al. 2012 and Tatarnikova et al. 2011. Squares are the new data.
               Times of our new spectroscopic observations are marked by vertical bars.
    }
   \label{Fig1}
   \end{figure}

Our {\it JHKL} photometry was performed with the 1.25-m ZTE telescope of the South Station of the Sternberg Astronomical Institute, Moscow State University, with the use of a single-channel InSb-photometer. The standard star was BS 7635 ({\it J} = 0.79, {\it H} = 0.03, {\it K} = −0.16, {\it L} =
−0.36). The uncertainties of the photometric measurements are within 0.03 mag.

Photometrical observations in {\it UBV} and {\it BVRI} systems were obtained on the two Zeiss-600 telescopes of the South Station, equipped with the photoelectric photometer of V.M. Lyuty and a CCD-photometer with FLI PL-4022 camera, respectively. We also use photometrical observations obtained at the Skalnat\'e Pleso Observatory of the Astronomical Institute of Slovak Academy of Sciences (see details in Shugarov et al. 2012). The star HD 192712 was a
UBV standard for photoelectric observations. For our CCD photometry we used three stars  "b", "c" and  "$\alpha$"  from  Henden \& Munari (2006) as main standard stars. The observational errors do not exceed 0.01 mag for all bands except $U$ band (in which the errors can reach 0.03 mag).

Spectroscopic observations were acquired with the 2.6-m G.A. Shain reflector (ZTSh) of the Crimean Astrophysical Observatory (CrAO) by using the SPEM slit spectrograph mounted at Nasmyth focus. The detector was a 1340*100 SPEC-10 CCD camera. The fixed 3 arcsec slit width allows spectra with a resolution of 8 \AA to be taken. The grating has 651 rulings/mm, yielding a dispersion of about 2 \AA/pixel. The primary reduction of the spectra, consisting of bias subtraction and flat-fielding, were done using the SPERED code developed by S.G. Sergeev at the CrAO. The subsequent reduction of the spectra, the wavelength calibration, and the spectral flux calibration were performed using the SPE code of Sergeev. The fitting of the spectra to a wavelength scale was done using the spectrum of a neon lamp. The calibration of the spectral fluxes was based on the absolute spectral energy distribution (SED) of the spectrophotometric standard HR 7744; the calibration uncertainty is within 10 \% (except the Balmer jump region, where uncertainty may be up to 30 \%) .

Spectroscopic observations of 2016 were obtained with the 1.25-m ZTE telescope by using a slit spectrograph equipped with a diffraction grating of 600 lines per mm and ST-402 CCD. The dispersion for these spectra is 2.3 \AA /pixel. The standard star was 40 Cyg and the calibration uncertainty is within 10\%. The primary reduction of the spectra were done using the ccdops. The subsequent reduction of these spectra were also performed using the SPE code.

\section{A light curves analysis}
\label{sect:LC}

As shown in Fig.1 the third eclipse differs greatly from the previous two ones . Minimum I and II are sharp and contact points are well-defined, while minimum III has got a sine-wave shape (with an amplitude in the {\it U} band of about 0.7 mag). Such type of light curves is typical for symbiotic stars in quiescent state. Minimum I (1980) was associated to a total eclipse of the hot component in supergiant stage (the main source of UV and optical radiation).

Therefore this minimum was deep and sharp and looked like a textbook eclipse in a binary system. However in the early 2000s the hot component of PU~Vul was in the hot subdwarf stage and its temperature was about 150000 K (see section "Spectral evolution"). Therefore in the third orbital cycle the main source of radiation in the near UV and blue ranges is an optically thin large gaseous nebula. The sine-wave shape of minimum III is related to the partial eclipse of the nebula by the orbiting cool component of PU~Vul.

The orbital brightness variations are clearly visible in {\it UBVR} bands where the nebula provides a significant part of the total system flux and they nearly disappear in {\it IJHKL} bands where the cool component radiation dominates.

   \begin{figure}
   \centering
   \includegraphics[width=10cm, angle=0]{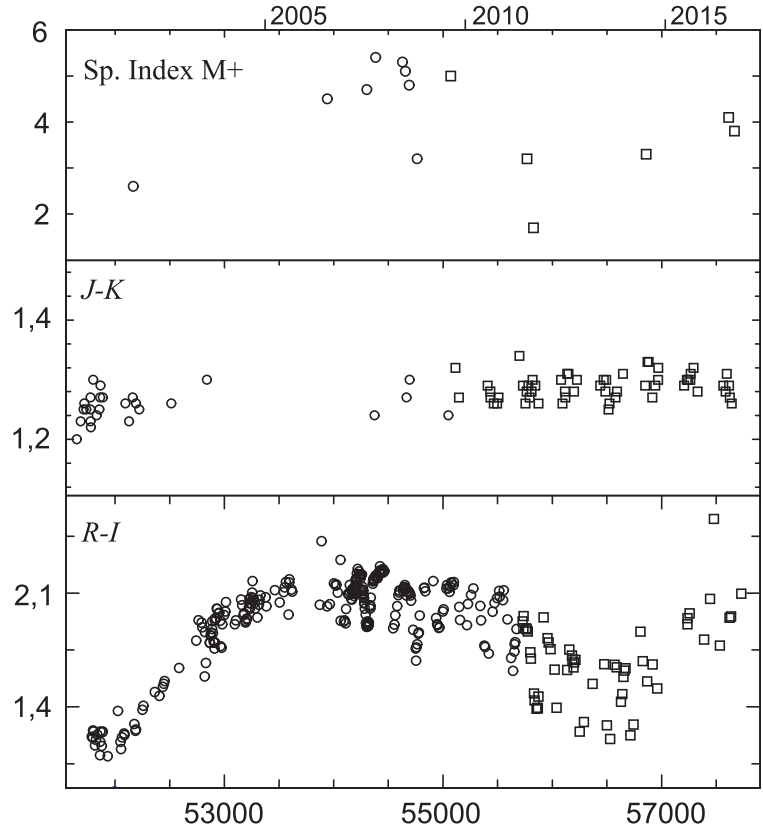}
   \caption{The R-I and J-K color curves and spectral class of the cool component.
                Circles are the data from Tatarnikova et al. 2011. Squares are the new data.
    }
   \label{Fig2}
   \end{figure}

   \begin{figure}
   \centering
   \includegraphics[width=\textwidth, angle=0]{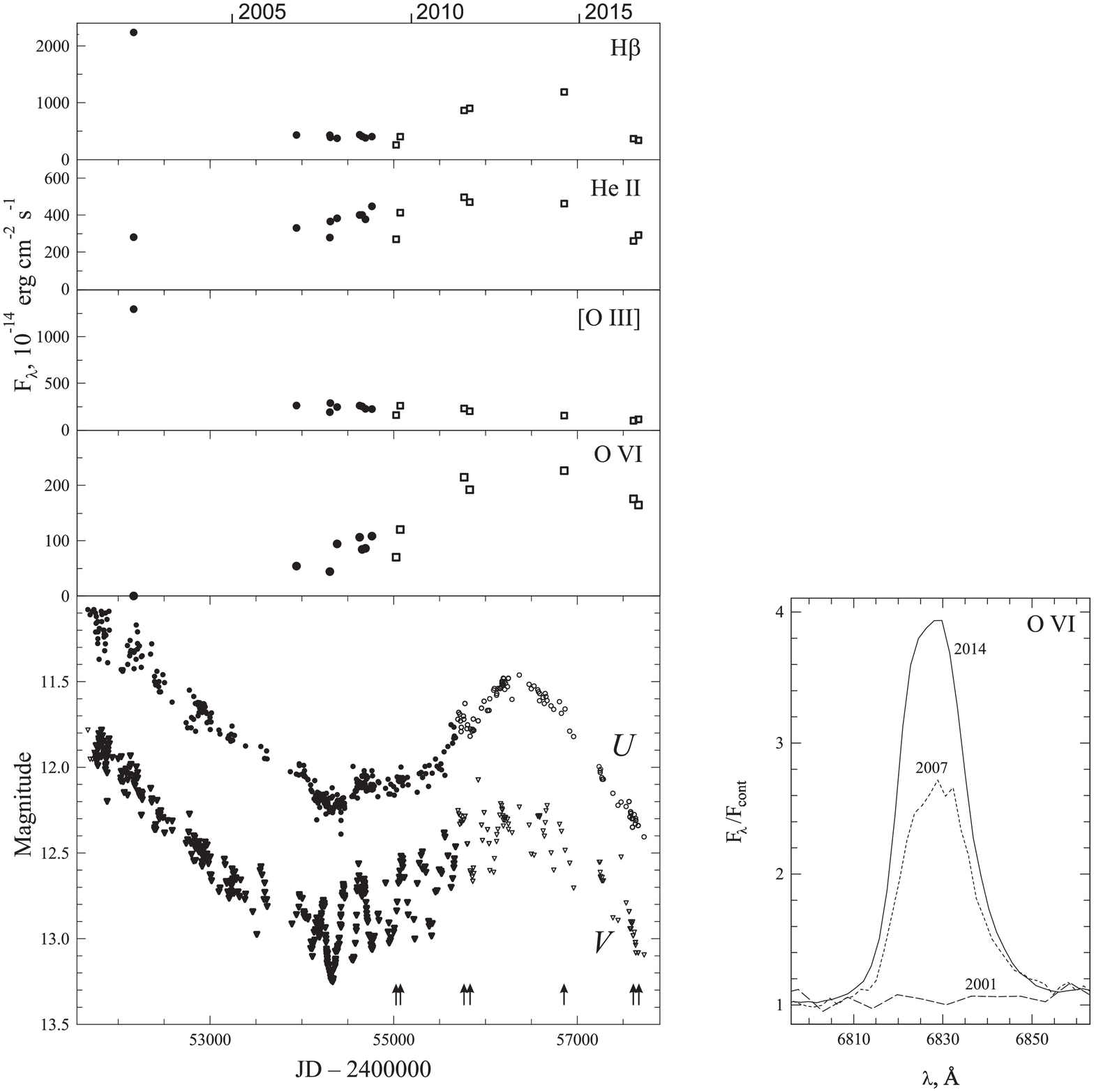}
   \caption{ Light curves of the symbiotic nova PU Vul for the {\it U}, {\it V} filters and some emission
   lines fluxes: H$\beta$, HeII (4686 {\AA}), [OIII] (5007 {\AA}) and OVI (6828{\AA}) -- left column (filled symbols are the data from Tatarnikova et al. 2011., open symbols are the new data, arrows mark the dates of our spectroscopic observations). The OVI line profile variations -- right column. 
    }
   \label{Fig3}
   \end{figure}

During the third orbital cycle the brightness variations interpreted as the cool component pulsations are clearly visible in the {\it VRIJHKL} light curves (Fig. 1). The period according to Shugarov et al. (2012) is about 217.7 days and the variations amplitude increases from 0.5 mag in {\it V} to 0.8 mag in {\it I}.  These variations become more prominent near eclipsing minimum III when the influence of the nebula radiation veiling  the cool component pulsations weakens sufficiently. Let's note that the R-I color index demonstrates a dependence that is not typical for red semiregular variables and miras: it is redder near pulsation minima (see Fig. 2). Whereas the J-K color index is barely dependent on pulsation phase. The reason is that the nebular radiation contributes to the total flux (including the strong H$\alpha$ emission) more essentially in the {\it R} band than in the {\it IJHK} bands. Consequently, the star, according to the R-I color index, is redder near pulsation maxima than it is near pulsation minima due to the relative contribution of the red giant emission to the {\it R} band being larger at those moments.

In addition to orbital brightness variations and cool component pulsations a long-term brightness evolution after the outburst of 1977 can be traced in Fig. 1.

\section{The spectral evolution}
\label{sect:SEv}

Variations of some optical lines fluxes are shown in Fig. 3. Only H$\beta$ emission line demonstrates strong orbital variations. Maybe He~II, 4686 {\AA} and Raman scattering line OVI, 6828 {\AA}  fluxes  are also correlated with orbital phase but very slow hot component evolution after the outburst (its temperature has increased since 1985) affects these emission line fluxes and distorts phase correlation. For example, two spectra (14.09.2001 (see Tatarnikova et al. 2011) and 18.07.2014) were obtained at nearly the same orbital phase ($\varphi \approx 0.56$ and $\varphi \approx 0.51$ accordingly, see ephemeris in Shugarov et al. 2012), however the emission feature OVI was absent in 2001 but it was one of the strongest emission lines in the spectrum in 2014.

The TiO 6180, 7100 {\AA} absorption bands are often used to estimate the spectral class of cool components of symbiotic stars. But all these estimates yield earlier spectral class for PU~Vul than the real  one is. The reason is that blue nebular continuum veils true depths of TiO bands. Fig. 2 gives evidence that red giant spectral class estimates are dependent not only on orbital phase but also on pulsation phase. The VO, 7865 {\AA} absorption band index provides a more reliable method for spectral class estimating due to the longer wavelength of the band. Only spectra obtained in 2016 include this band. The cool component spectral class estimated from VO is about M5.7-M6. A lack of spectral data doesn't allow us to investigate possible real variability of the cool component spectral class. We used M5III and M6III standard SEDs to model total continuum emission during the whole period of spectral observations. The model SEDs  for red giants of various spectral types were taken from the spectral library by Silva and Cornell 1992.

Fig.4 and 5 display spectra of PU~Vul deredenned with E(B-V)=0.4 mag (Vogel \& Nussbaumer 1992) and appropriate model SEDs consisting of the hot component radiation, the optically thin radiation from the nebula, which absorbs all the Lc-photons and the radiation from a standard red giant (see details in Esipov et al. 2000). The adopted electron temperature of the nebula was about 20000~K. The hot component temperature was estimated by means of modified Zanstra method (see Tatarnikova et al. 2011) applied to the modified equivalent width of the He~II, 4686 {\AA} line (the flux in the He~II emission line was referred to the flux in the blue continuum near 3600 {\AA}). We determined the hot component luminosity from its bolometric flux (we estimated it during modelling the SEDs) and distance to the system (D=3.5 kpc, Tatarnikova \& Tatarnikov 2009). Note that our previous hot component temperature estimate (see Tatarnikova et al. 2011) is not correct because it was obtained in 2008 when the sufficient part of the nebula was eclipsed by the cool component. But by that moment our photometrical observations had covered only half of the third orbital cycle and therefore the light curves looked like a gradual brightness decrease, not like an eclipse III. New spectral observations allow us to draw another point in the temperature-luminosity diagram for PU~Vul in 2014 when the effect of the cool component eclipsing the nebula was negligible (see Fig. 6).

\begin{figure}[h]
  \begin{minipage}[t]{0.495\linewidth}
  \centering
   \includegraphics[width=70mm]{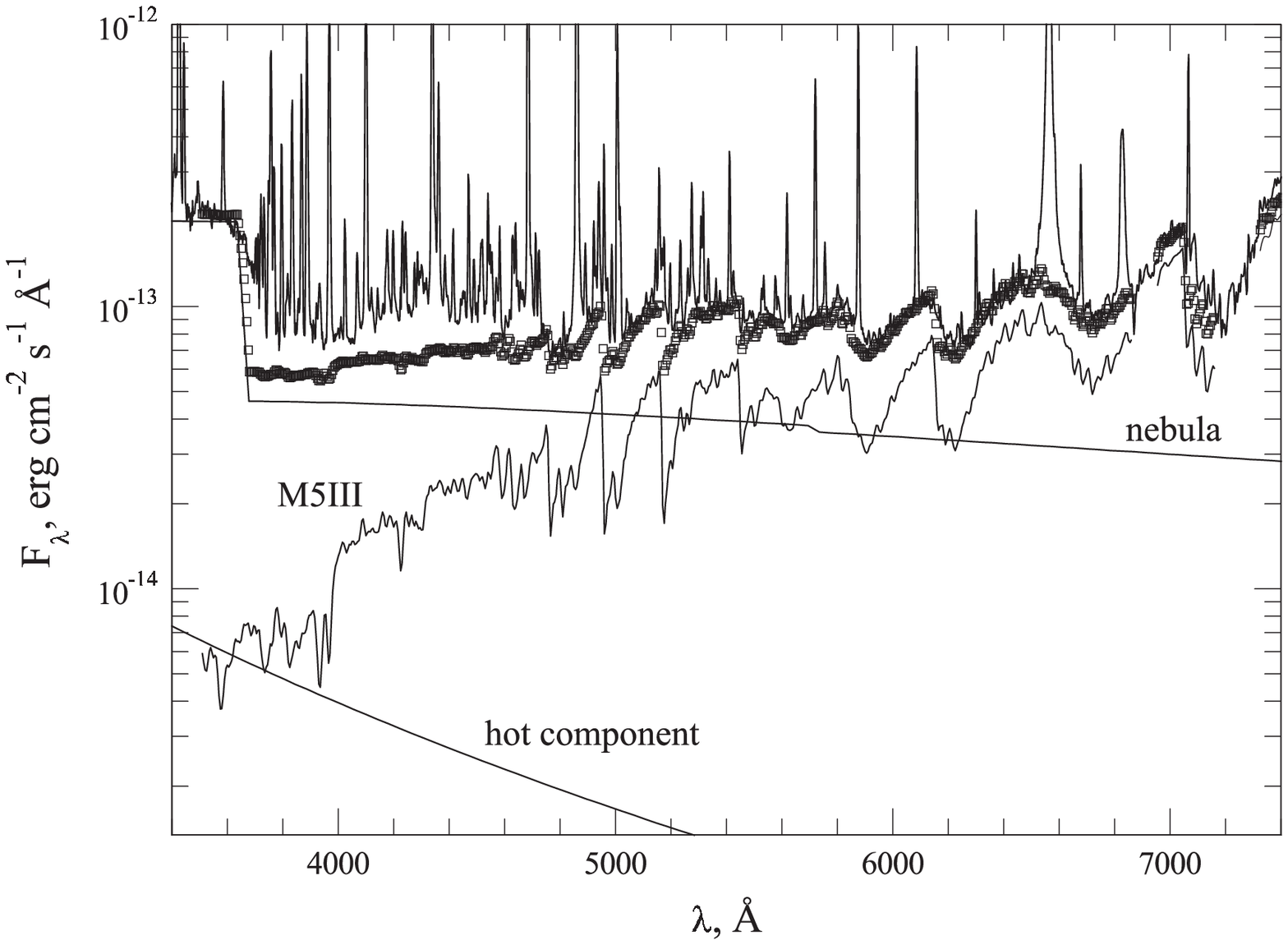}
   \captionsetup{margin=5pt}
   \caption{{\small Interstellar-reddening-corrected spectrum of PU~Vul 
                obtained on July 18, 2014 (solid line). The squares denote the model continuum SED consisting of 
                light from  hot component ($T_{hot} = 147000 K$), optically thin nebula 
               ($T_e = 20000 K$) that absorbs all Lc-quanta and the cool component (spectral type M5III).
    } }
  \end{minipage}%
  \begin{minipage}[t]{0.495\textwidth}
   \centering
   \includegraphics[width=70mm]{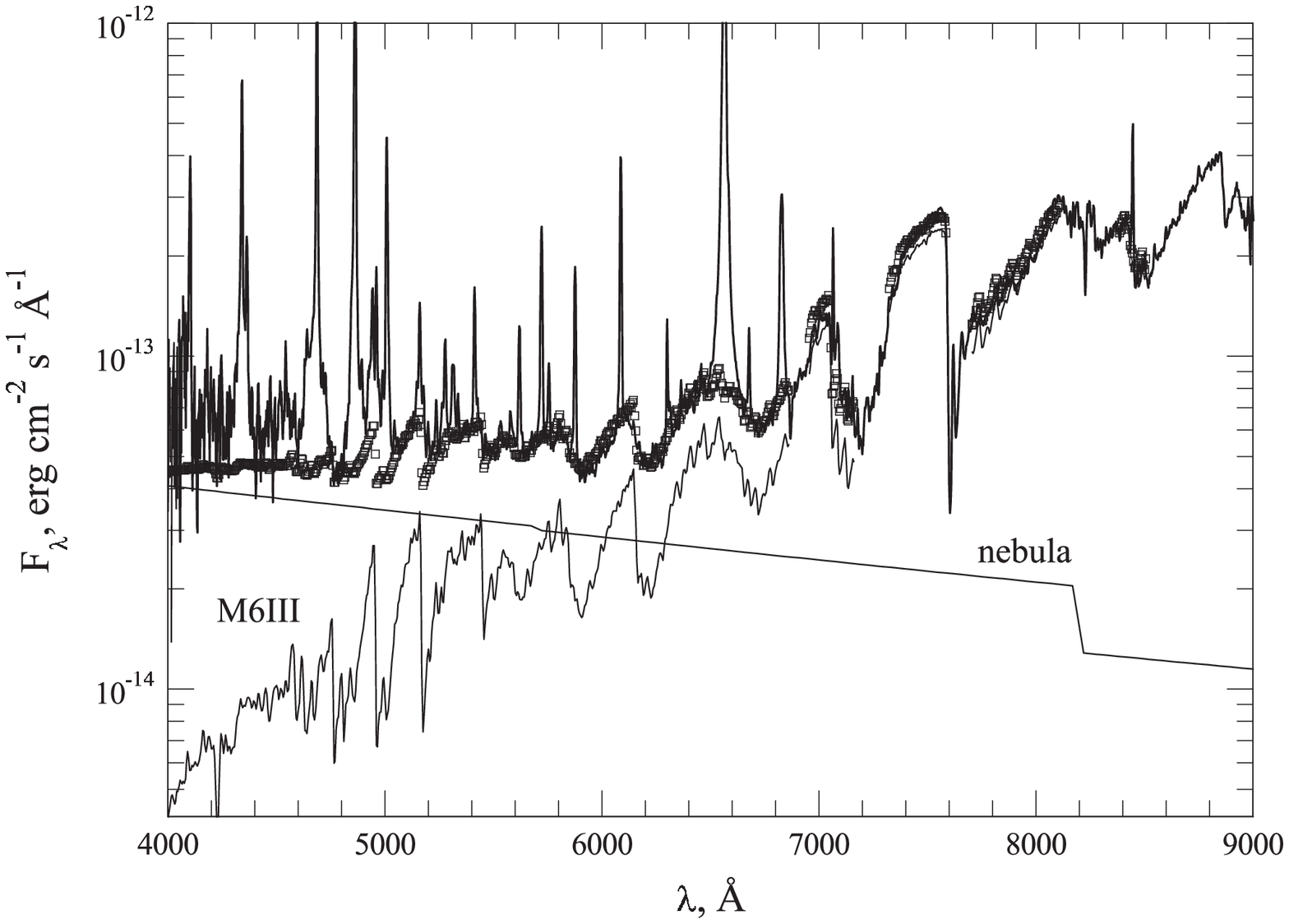}
   \captionsetup{margin=5pt}
   \caption{{\small Interstellar-reddening-corrected spectrum of PU~Vul  obtained on August 11, 2016 (solid line). 
                 The squares denote the model continuum SED consisting of light from  the cool component (spectral type M6III) and optically thin nebula ($T_e = 20000 K$) 
    }}
  \end{minipage}%
  \label{Fig:fig4-5}
\end{figure}

   \begin{figure}
   \centering
   \includegraphics[width=10cm, angle=0]{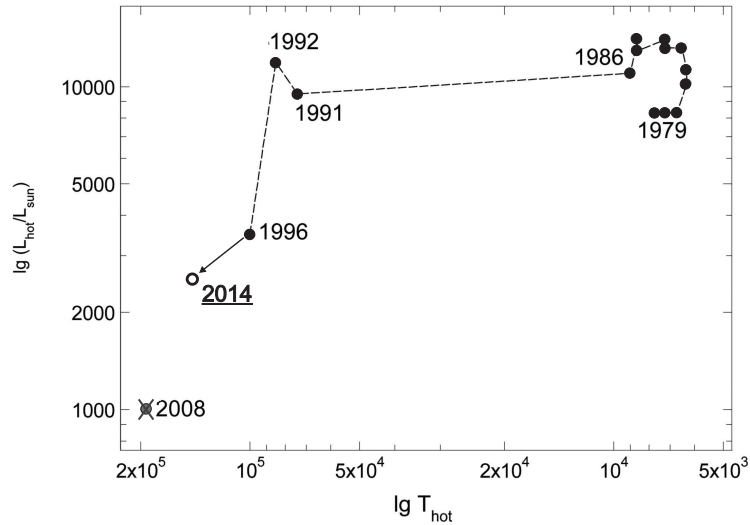}
   \caption{Positions of the hot component of PU Vul in the Hertzsprung–Russell diagram in 1979–2014. 
            The star’s positions in 1979–1996 were taken from Tatarnikova \& Tatarnikov, 2009 (see also references therein).
    }
   \label{Fig6}
   \end{figure}

\section{Conclusions}
\label{sect:conclusion}

The analysis of our spectral and photometric observations of PU~Vul reveals the presence of three types of variability: 1) the variability related to orbital motion, 2) the cool component pulsations, 3) the brightness variations due to slow after-outburst evolution of the hot component.  

During the third orbital cycle the UBVR light curves demonstrated a sine-wave shape which is typical for symbiotic stars in quiescent state.

The hot component temperature is still gradually increasing whereas its luminosity is simultaneously decreasing. Therefore the hot component of PU~Vul is not on the cooling curve yet.

\begin{acknowledgements}
The effort of A.A. Tatarnikova is  supported  by  RNF  grant  17-12-01241 (problem  formulation and the analysis). S. Shugarov thanks for partial support the grants VEGA 2/0008/17 and APPV 15-0458. A. M. Tatarnikov thanks for partial support the grant RFBR 15-07-04512-A.
\end{acknowledgements}

\label{lastpage}

\end{document}